\documentclass[10pt]{article}
\usepackage[utf8]{inputenc}
\usepackage{graphicx}
\usepackage{amsmath}
\usepackage{amsfonts}
\usepackage{amssymb}
\usepackage{physics}
\usepackage{siunitx}

\usepackage{epstopdf}
\AppendGraphicsExtensions{.tif}

\usepackage{setspace}
\usepackage{authblk}
\usepackage{caption}
\usepackage{float}
\usepackage{cite}
\usepackage[left=2.0cm,right=2.0cm,top=2cm,bottom=2cm]{geometry}
\usepackage{color}
\usepackage{color,soul}
\usepackage{multicol}
\usepackage{subcaption}
\usepackage{hyperref}

\title{\textbf{Spatiotemporal Stabilization of Turbulence-Distorted Gaussian Beams via Waveguide Spatial Filtering}}

\author[1,*]{Shouvik Sadhukhan}
\author[4]{C. S. Narayanamurthy}

\affil[1, 4]{\small{Applied and Adaptive Optics Laboratory, Department of Physics, Indian Institute of Space Science and Technology (IIST), P.O: Valiamala, Trivandrum - 695547, State: Kerala; India}}
\affil[1]{\small{Email: shouvikphysics1996@gmail.com}}
\affil[4]{\small{Email: naamu.s@gmail.com}}
\affil[*]{\small{Corresponding Author Email: shouvikphysics1996@gmail.com}}

\begin{document}
\maketitle

\begin{abstract}
Optical beams propagating through atmospheric turbulence undergo spatiotemporal intensity fluctuations that deviate significantly from an ideal Gaussian profile. In this work, we present a unified theoretical and experimental framework for quantifying and mitigating these turbulence-induced distortions by coupling a higher-order statistical characterization technique with optical waveguide spatial filtering. The statistical characterization employs a Cholesky-whitened Gram--Charlier expansion that decomposes the two-dimensional beam intensity distribution into a Gaussian core augmented by third- and fourth-order cumulant corrections, thereby isolating skewness and excess kurtosis as quantitative non-Gaussianity indicators. Concurrently, the propagation of the distorted beam through a dielectric waveguide is analyzed to demonstrate that higher-order spatial modes, which carry the dominant share of turbulence-induced structural distortions, encounter a cutoff condition governed by the normalized frequency parameter and subsequently undergo exponential attenuation along the propagation direction. The waveguide thus acts as a passive spatial mode filter that selectively transmits the fundamental guided mode while suppressing radiative higher-order modes. The fitted beam volume, derived from the Gram--Charlier intensity model, serves as a unified scalar diagnostic that tracks the frame-by-frame evolution of turbulence-induced structural changes. Experimental measurements validate the theoretical predictions, demonstrating a substantial reduction in intensity fluctuations and a recovery of Gaussian beam statistics after waveguide propagation.\\

\textbf{Keywords:} Atmospheric turbulence, Gaussian beam, Gram--Charlier expansion, waveguide spatial filtering, beam intensity stabilization, spatiotemporal fluctuations.
\end{abstract}

\section{Introduction}\label{sec:intro}

Laser beam propagation through atmospheric turbulence is a problem of central importance in free-space optical communications, laser ranging, directed energy, and adaptive optics systems \cite{62}. The refractive index fluctuations arising from thermally driven turbulent eddies impose random phase perturbations on a propagating wavefront, leading to beam wander, beam broadening, and intensity scintillation at the receiver plane. These effects collectively degrade the spatial coherence and pointing stability of the beam, and their severity increases with turbulence strength and propagation distance.

Classical descriptions of turbulence-induced beam degradation rely on second-order statistical quantities such as the mean beam width, centroid variance, and scintillation index \cite{2,5}. While these parameters provide useful global measures of beam quality, they are insufficient to characterize the full morphology of the distorted intensity profile, particularly under strong turbulence conditions where the distribution departs significantly from a Gaussian envelope. Higher-order statistical features including skewness and excess kurtosis encode asymmetric distortions and heavy-tailed structures in the beam profile that are invisible to second-order analysis alone.

On the mitigation side, adaptive optics systems represent the most established approach to correcting wavefront aberrations in real time \cite{62}. However, their complexity, cost, and finite temporal bandwidth limit their applicability in many practical scenarios. Passive mitigation strategies, including aperture averaging, beam shaping, and spatial filtering, offer complementary approaches that do not require active wavefront sensing. Among these, the propagation of a distorted beam through a single-mode or few-mode optical waveguide offers a particularly elegant mechanism: the waveguide geometry imposes a discrete modal structure on the transmitted field, and modes that do not satisfy the confinement condition are exponentially attenuated along the propagation direction, leaving only the supported guided modes to emerge at the output.

Despite the long-standing recognition of waveguides as spatial mode filters in integrated photonics and fiber optic systems, their explicit role in suppressing turbulence-induced higher-order beam distortions has not been analyzed within a rigorous higher-order statistical framework. In particular, the connection between the cumulant-based non-Gaussianity indicators of the input beam and the modal attenuation characteristics of the waveguide has not been established in a quantitative manner.

In this work, we develop a comprehensive theoretical framework that bridges these two aspects. First, we formulate a statistical model for the two-dimensional beam intensity distribution based on a normalized probability measure. After computing the beam centroid and covariance matrix, a Cholesky whitening transformation \cite{1} is applied to standardize the spatial coordinates, isolating genuine non-Gaussian features in the form of third- and fourth-order cumulants. These cumulants are incorporated into a bivariate Gram--Charlier expansion that provides a closed-form analytical representation of the distorted beam profile. Second, we present a derivation of the cutoff condition and modal decay characteristics of a symmetric slab waveguide. Starting from the Helmholtz equation, we show that higher-order modes---which are directly responsible for the non-Gaussian distortions captured by the cumulant analysis---become radiative when they exceed the cutoff condition, and subsequently attenuate exponentially during propagation. Together, these two analyses demonstrate that waveguide propagation acts as a natural mechanism for recovering Gaussian beam statistics by physically suppressing the modes responsible for skewness and kurtosis excess.

The remainder of the paper is organized as follows. Section~\ref{sec:theory} presents the theoretical framework, including the probabilistic intensity model, the Gram--Charlier cumulant expansion, and the waveguide modal analysis. The experimental setup is described in Section~\ref{sec:experiment}. Results and discussion are presented in Section~\ref{sec:results}, and conclusions are drawn in Section~\ref{sec:conclusion}. Detailed derivations are collected in the Appendix.

\section{Theoretical Framework}\label{sec:theory}

\subsection{Statistical Characterization of the Turbulence-Distorted Beam}

\subsubsection{Probability Measure from Intensity Distribution}

Let $I_{ij} \geq 0$ denote the measured two-dimensional intensity on a pixel grid with spatial coordinates $(x_j, y_i)$. We interpret the normalized intensity as a discrete probability measure,
\begin{equation}
    p_{ij} = \frac{W_{ij}}{S}, \quad S = \sum_{i,j} W_{ij},
    \label{eq:prob_measure}
\end{equation}
where $W_{ij}$ are weighted intensities (incorporating any masking). For any spatial function $f(X,Y)$, the intensity-weighted expectation is $\mathbb{E}[f] = \sum_{i,j} p_{ij} f(x_j, y_i)$.

\subsubsection{Centroid, Covariance, and Cholesky Whitening}

The beam centroid and spatial covariance are,
\begin{equation}
    \mu_x = \mathbb{E}[X], \quad \mu_y = \mathbb{E}[Y], \quad
    \Sigma = \begin{pmatrix} \sigma_{xx} & \sigma_{xy} \\ \sigma_{xy} & \sigma_{yy} \end{pmatrix}.
    \label{eq:covariance}
\end{equation}
Since $\Sigma$ is symmetric positive definite, it admits a unique lower-triangular Cholesky factorization $\Sigma = LL^\top$. The whitened coordinates,
\begin{equation}
    \mathbf{z} = L^{-1} \begin{pmatrix} x - \mu_x \\ y - \mu_y \end{pmatrix},
    \label{eq:whitening}
\end{equation}
satisfy $\mathbb{E}[\mathbf{z}] = \mathbf{0}$ and $\mathrm{Cov}(\mathbf{z}) = I_2$, ensuring that all second-order structure is normalized and any residual non-Gaussianity is isolated in higher-order statistics.

\subsubsection{Cumulants and Non-Gaussianity Measures}

Within the whitened frame, the standardized moments $m_{pq} = \sum_k w_k z_{1k}^p z_{2k}^q$ are computed. The whitening constraints enforce $m_{20} = m_{02} = 1$ and $m_{11} = 0$. The third- and fourth-order cumulants are (see Appendix~\ref{app:cumulants} for full relations),
\begin{align}
    &k_{30} = m_{30},\quad k_{21} = m_{21},\quad k_{12} = m_{12},\quad k_{03} = m_{03}, \label{eq:skew_cum}\\
    &k_{40} = m_{40} - 3,\quad k_{04} = m_{04} - 3,\quad k_{22} = m_{22} - 1. \label{eq:kurt_cum}
\end{align}
The overall degree of non-Gaussianity is quantified by the Euclidean norms,
\begin{equation}
    |\mathrm{skew}|_3 = \sqrt{k_{30}^2 + k_{21}^2 + k_{12}^2 + k_{03}^2}, \quad
    |\mathrm{kurt}|_4 = \sqrt{k_{40}^2 + k_{31}^2 + k_{22}^2 + k_{13}^2 + k_{04}^2}.
    \label{eq:norms}
\end{equation}

\subsubsection{Gram--Charlier Intensity Model}

The distorted beam intensity is represented as a Gaussian core modulated by Hermite polynomial corrections. In the whitened coordinate system, the Gram--Charlier expansion reads,
\begin{equation}
    p(\mathbf{r}) \approx \phi(\mathbf{r})\left[1 + \frac{1}{6}\left(k_{30}H_{30} + 3k_{21}H_{21} + 3k_{12}H_{12} + k_{03}H_{03}\right) + \frac{1}{24}\left(k_{40}H_{40} + 4k_{31}H_{31} + 6k_{22}H_{22} + 4k_{13}H_{13} + k_{04}H_{04}\right)\right],
    \label{eq:gc_expansion}
\end{equation}
where $\phi(\mathbf{r}) = (2\pi\sqrt{\det\Sigma})^{-1}\exp\!\left[-\tfrac{1}{2}(\mathbf{r}-\boldsymbol{\mu})^\top\Sigma^{-1}(\mathbf{r}-\boldsymbol{\mu})\right]$ is the reference Gaussian, and $H_{\alpha_1\alpha_2}(\mathbf{z})$ are bivariate Hermite polynomials (listed in Appendix~\ref{app:hermite}). The fitted intensity is,
\begin{equation}
    I_{\mathrm{fit}}(\mathbf{r}) = a\, p(\mathbf{r}) + b,
    \label{eq:fit}
\end{equation}
with amplitude $a$ and background $b$ determined by least-squares fitting.

\subsubsection{Fitted Power Volume as a Turbulence Diagnostic}

Dynamic turbulence displaces the beam centroid and alters the covariance and cumulants on a frame-by-frame basis. To provide a unified scalar diagnostic, we define the fitted power volume,
\begin{equation}
    V_{\mathrm{frame}} = \int_{\mathbb{R}^2} I_{\mathrm{fit}}(\mathbf{r},t)\,\mathrm{d}^2r = 2\pi I_{\mathrm{max}}\sqrt{|\Sigma|}.
    \label{eq:volume}
\end{equation}
Although $V_{\mathrm{frame}}$ does not represent the total optical power (since the expansion is truncated at fourth order), its frame-to-frame variation directly encodes turbulence-induced structural changes and therefore serves as a robust indicator of beam distortion severity.

\subsection{Waveguide Spatial Mode Filtering}

\subsubsection{Modal Structure of a Dielectric Slab Waveguide}

Consider a symmetric slab waveguide with core refractive index $n_1$ and cladding index $n_2 < n_1$, core half-width $d/2$. The transverse field distribution satisfies the Helmholtz eigenvalue equation,
\begin{equation}
    \frac{\mathrm{d}^2\psi}{\mathrm{d}x^2} + \left(k^2 n^2 - \beta^2\right)\psi = 0,
    \label{eq:helmholtz_t}
\end{equation}
where $k = \omega/c$ and $\beta$ is the longitudinal propagation constant. Guided modes require $kn_2 < \beta < kn_1$, yielding oscillatory fields inside the core and exponentially decaying fields in the cladding ($\psi \propto e^{-\gamma|x|}$, $\gamma^2 = \beta^2 - k^2n_2^2 > 0$). Applying continuity boundary conditions at the core--cladding interface produces the modal eigenvalue equations,
\begin{equation}
    \tan\!\left(\frac{k_x d}{2}\right) = \frac{\gamma}{k_x} \quad \text{(even modes)}, \qquad
    -\cot\!\left(\frac{k_x d}{2}\right) = \frac{\gamma}{k_x} \quad \text{(odd modes)},
    \label{eq:eigenvalue_eq}
\end{equation}
with $k_x^2 = k^2 n_1^2 - \beta^2$.

\subsubsection{Normalized Frequency and Mode Count}

Introducing the normalized frequency (V-number),
\begin{equation}
    V = kd\sqrt{n_1^2 - n_2^2},
    \label{eq:vnumber}
\end{equation}
the maximum number of supported guided modes is,
\begin{equation}
    m_{\max} \approx \frac{V}{\pi}.
    \label{eq:mmax}
\end{equation}
By engineering $V$ to satisfy $V < \pi$ (single-mode condition) or a small integer multiple thereof, the waveguide can be designed to transmit only the fundamental mode, thereby rejecting the higher-order spatial modes that carry turbulence-induced distortions.

\subsubsection{Cutoff Condition and Exponential Decay of Higher-Order Modes}

A mode of order $m$ reaches cutoff when $\gamma \to 0$, i.e., $\beta \to kn_2$. Beyond cutoff the propagation constant satisfies $\beta < kn_2$, becoming imaginary ($\beta_m = i\alpha$). The electric field then decays exponentially,
\begin{equation}
    E(z) = E_0\,e^{i\beta_r z}\,e^{-\alpha_z z},
    \label{eq:field_decay}
\end{equation}
and the intensity attenuates as,
\begin{equation}
    I(z) = I_0\,e^{-2\alpha_z z}.
    \label{eq:intensity_decay}
\end{equation}
For a waveguide without cladding but with a finite transverse aperture $d$, the same conclusion follows from the boundary condition $\psi(\pm d/2)=0$, which quantizes the transverse wave number as $k_t = m\pi/d$ and yields,
\begin{equation}
    \beta_m^2 = k^2 n^2 - \left(\frac{m\pi}{d}\right)^2.
    \label{eq:beta_m}
\end{equation}
Modes for which $\beta_m^2 < 0$ are evanescent and decay as $e^{-\alpha z}$ with $\alpha = \sqrt{(m\pi/d)^2 - k^2n^2}$ (see Appendix~\ref{app:waveguide} for full derivation). Physically, higher-order modes correspond to larger transverse momenta and therefore larger ray angles $\theta$ relative to the waveguide axis. When $\theta$ exceeds the critical angle $\theta_c = \arcsin(n_2/n_1)$, the mode refracts into the cladding and is lost, leaving only the guided modes to propagate.

\subsubsection{Connection to Non-Gaussian Beam Distortions}

The non-Gaussian cumulants $k_{30},\, k_{21},\, k_{12},\, k_{03}$ (skewness) and $k_{40},\, k_{31},\, k_{22},\, k_{13},\, k_{04}$ (excess kurtosis) are sensitive to the higher spatial frequency content of the beam profile. These contributions arise precisely from the higher-order spatial modes of the field. As the beam is coupled into the waveguide and the higher-order modes are exponentially attenuated according to Eq.~\eqref{eq:intensity_decay}, the residual transmitted field approaches the fundamental Gaussian mode, and the cumulant norms $|\mathrm{skew}|_3$ and $|\mathrm{kurt}|_4$ decrease toward zero. Thus, the waveguide acts as a passive, all-optical mechanism for recovering near-Gaussian beam statistics, and the fitted volume $V_{\mathrm{frame}}$ is expected to stabilize toward a constant value as turbulence-induced higher-order modes are filtered out.

\section{Scintillation Suppression with Spatial Mode Filtering}

Atmospheric turbulence introduces random phase perturbations into propagating optical beams, redistributing optical energy among multiple spatial modes and producing strong intensity fluctuations at the observation plane. When such a distorted beam is coupled into a dielectric waveguide or optical fiber, the guided structure imposes a discrete modal basis on the field. Modes that satisfy the guiding condition propagate, while higher-order modes exceeding the cutoff condition become evanescent and decay exponentially along the propagation direction. This process naturally suppresses modal interference and reduces the scintillation of the transmitted beam.

\subsection{Modal Decomposition of the Turbulence-Distorted Field}

Let the complex electric field of a turbulence-distorted beam at the waveguide input plane $z=0$ be

\begin{equation}
E(x,y,0)=A(x,y)e^{i\phi(x,y)},
\end{equation}

where $A(x,y)$ denotes the amplitude envelope and $\phi(x,y)$ represents turbulence-induced phase distortions. Because the waveguide supports a set of orthonormal transverse eigenmodes $\psi_m(x,y)$, the field can be expanded as

\begin{equation}
E(x,y,0)=\sum_{m} a_m \psi_m(x,y),
\end{equation}

where the modal excitation coefficients are obtained through

\begin{equation}
a_m=\iint E(x,y,0)\psi_m^{*}(x,y)\,dx\,dy.
\end{equation}

Atmospheric turbulence generally excites a broad spectrum of spatial modes, particularly higher-order modes that correspond to high spatial-frequency distortions of the beam profile.

\subsection{Propagation of Waveguide Modes}

Inside the waveguide each mode propagates independently with propagation constant $\beta_m$, yielding

\begin{equation}
E(x,y,z)=\sum_{m} a_m \psi_m(x,y)e^{i\beta_m z}.
\end{equation}

The resulting intensity distribution is

\begin{equation}
I(x,y,z)=|E(x,y,z)|^2 .
\end{equation}

Expanding the squared modulus gives

\begin{equation}
I(x,y,z)=\sum_{m}|a_m|^2 |\psi_m|^2
+
\sum_{m\neq n} a_m a_n^* \psi_m \psi_n^* e^{i(\beta_m-\beta_n)z}.
\end{equation}

The second summation represents modal interference terms responsible for spatial and temporal intensity fluctuations.

\subsection{Cutoff and Evanescent Decay of Higher-Order Modes}

The transverse modal structure of the waveguide follows from the Helmholtz equation

\begin{equation}
\nabla^2 E + k^2 n^2 E =0.
\end{equation}

Assuming a modal solution of the form

\begin{equation}
E(x,z)=\psi(x)e^{i\beta z},
\end{equation}

one obtains the transverse eigenvalue equation

\begin{equation}
\frac{d^2\psi}{dx^2}+(k^2 n^2-\beta^2)\psi=0.
\end{equation}

Guided modes require the propagation constant to satisfy

\begin{equation}
k n_2 < \beta_m < k n_1 ,
\end{equation}

where $n_1$ and $n_2$ are the refractive indices of the core and cladding respectively. When this condition is violated ($\beta_m < k n_2$), the propagation constant becomes imaginary,

\begin{equation}
\beta_m = i\alpha_m ,
\end{equation}

and the corresponding field becomes evanescent,

\begin{equation}
E_m(x,y,z)=a_m\psi_m(x,y)e^{-\alpha_m z}.
\end{equation}

The modal intensity therefore decays exponentially,

\begin{equation}
I_m(z)=|a_m|^2 e^{-2\alpha_m z}.
\end{equation}

\subsection{Reduction of Scintillation}

The scintillation index, which quantifies intensity fluctuations, is defined as

\begin{equation}
\sigma_I^2=\frac{\langle I^2\rangle-\langle I\rangle^2}{\langle I\rangle^2}.
\end{equation}

Since the dominant contribution to the variance arises from modal interference, the fluctuation component evolves as

\begin{equation}
I_{\text{fluct}}(z)\propto
\sum_{m\neq n}|a_m(0)|^2|a_n(0)|^2
e^{-2(\alpha_m+\alpha_n)z}.
\end{equation}

Because higher-order modes exhibit larger decay constants $\alpha_m$, their amplitudes rapidly decrease during propagation. For sufficiently large propagation distances the field approaches the fundamental mode,

\begin{equation}
E(x,y,z)\approx a_0 \psi_0(x,y)e^{i\beta_0 z},
\end{equation}

leading to a stable output intensity

\begin{equation}
I(x,y,z)=|a_0|^2|\psi_0(x,y)|^2.
\end{equation}

Consequently, the scintillation index decreases with propagation distance as

\begin{equation}
\sigma_I^2(z)\propto
\sum_{m\neq n}|a_m|^2|a_n|^2
e^{-2(\alpha_m+\alpha_n)z},
\end{equation}

demonstrating that waveguide propagation acts as a passive spatial filter that suppresses turbulence-induced higher-order modes and restores a stable near-Gaussian beam profile.

\section{Experimental Setup}\label{sec:experiment}
\subsection{Optical System Configuration}

\begin{figure}[H]
\centering
\includegraphics[width=1.0\textwidth]{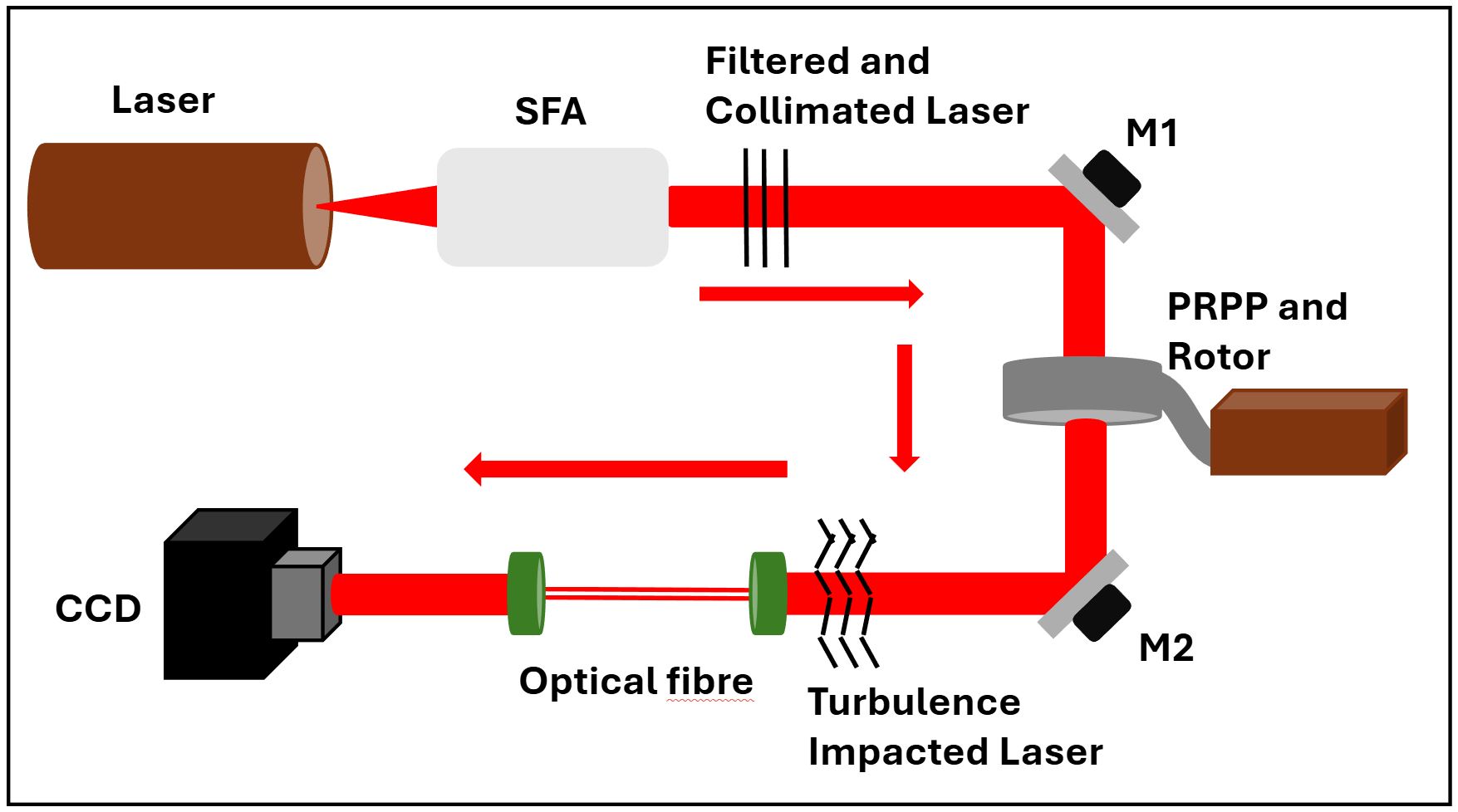}
\caption{Schematic of the experimental optical setup. A He-Ne laser beam is spatially filtered and collimated (SFA), redirected by mirror M1, passed through the rotating PRPP to introduce Kolmogorov turbulence, redirected by mirror M2, and coupled into an optical fibre waveguide before detection on a CCD camera. The PRPP rotor provides temporal variation of the turbulence realizations across the 200 frames captured per experimental set.}
\label{fig:setup}
\end{figure}

The experimental apparatus is illustrated in Figure~\ref{fig:setup}. A continuous-wave He-Ne laser operating at $\lambda = 632.8$~nm served as the coherent light source. The output beam was spatially filtered and collimated using a spatial filter assembly (SFA) consisting of a microscope objective, a pinhole, and a collimating lens, yielding a clean, diffraction-limited Gaussian beam of approximately 5~mm diameter ($1/e^2$). The collimated beam was redirected by a planar mirror (M1) and directed onto a Pseudo-Random Phase Plate (PRPP, Thorlabs EDU-RPP1) mounted on a motorized rotation stage, which introduced controlled Kolmogorov-type wavefront aberrations to simulate atmospheric turbulence (Section~\ref{sec:experiment}.2). After passing through the PRPP, the turbulence-impacted beam was redirected by a second planar mirror (M2) and coupled into a step-index optical fibre, which served as a passive waveguide compensation element. The transmitted beam emerging from the fibre output was subsequently imaged onto a CCD camera (Thorlabs DCC1545M) using an imaging lens, and the recorded intensity distributions were stored for offline analysis.

The optical fibre functioned as a spatial mode filter, selectively attenuating higher-order spatial modes introduced by the turbulence-induced wavefront distortions, thereby partially restoring the beam's spatial coherence at the detection plane. No adaptive-optics, wavefront-correction, or polarization elements were employed, making the system a purely passive scintillation-mitigation configuration based on guided-wave propagation. The turbulence-free reference dataset (Set~2) was acquired by removing the PRPP from the beam path while keeping all other components, including the optical fibre, in place, thereby establishing a baseline intensity distribution for the undisturbed Gaussian beam.

\subsection{Turbulence Simulation}
Atmospheric turbulence was simulated using a Pseudo-Random Phase Plate (PRPP, Thorlabs EDU-RPP1) consisting of five layers: two BK7 glass windows enclosing a central acrylic layer with imprinted Kolmogorov-type phase profile, stabilized by near-index-matching polymer layers. The plate generates aberrated wavefronts with adjustable Fried coherence lengths $r_0 \approx 0.6$ mm where $16$-$32$ samples distributed over 4096 phase points. The PRPP was mounted on a motorized rotation stage (rotation speed: 1 rpm) to simulate temporal turbulence dynamics.

\textit{Turbulence Scale Equivalence:} In laboratory turbulence simulations such as PRPP experiments, a Fried coherence length as small as $r_0 = 0.6\,\text{mm}$ does not represent natural atmospheric conditions but rather an artificially scaled turbulence strength designed to emulate long-distance atmospheric propagation within a short experimental path. The Fried parameter is given by
\[
r_0 = \left[0.423\, k_0^2 \int_0^L C_n^2(z)\, dz \right]^{-3/5},
\]
and for uniform turbulence simplifies to $r_0 = [0.423\, k_0^2 C_n^2 L]^{-3/5}$. Inverting this expression for $r_0 = 0.6\,\text{mm}$ at $\lambda = 632.8\,\text{nm}$ and $L = 1.5\,\text{cm}$ yields $C_n^2 \approx 3.35\times10^{-7}\,\text{m}^{-2/3}$, which is many orders of magnitude stronger than typical atmospheric values ($10^{-17}$--$10^{-13}\,\text{m}^{-2/3}$). This enhanced $C_n^2$ compensates for the short laboratory propagation length, effectively compressing kilometer-scale atmospheric turbulence into a meter-scale setup. Consequently, the small $r_0$ value reflects a strong turbulence regime used for controlled experimental emulation rather than natural atmospheric coherence conditions. The present scenario provides an equivalent atmospheric turbulence with a minimum 560 km propagation with the same Fried coherence length as PRPP. Detailed derivation has been provided in the supplementary file.

\subsection{Optical Fibre Waveguide}
A step-index optical fibre served as the passive compensation element, exploiting guided-wave propagation to mitigate turbulence-induced spatial intensity fluctuations. Key properties:

\begin{itemize}
    \item Core/cladding refractive indices: $n_\text{core} > n_\text{cladding}$ (step-index profile)
    \item Operating wavelength: $\lambda = 632.8$~nm
    \item Numerical aperture (NA): sufficient to couple the aberrated beam
    \item Transmission: $T > 90\%$ at $\lambda = 632.8$~nm
    \item Function: Spatial mode filtering via guided propagation
\end{itemize}

The optical fibre was selected for the following reasons: (1) High optical transmission with low absorption and scattering in the visible range; (2) Effective spatial mode filtering that suppresses higher-order turbulence-induced modes; (3) Sufficient guided propagation length for coherence restoration effects; (4) Compatibility with free-space beam coupling via standard fibre-coupling optics.

\subsection{Detection System}
A 8-bit CCD camera (Thorlabs DCC1545M, pixel size: 5.2~$\mu$m, sensor size: $1280\times1024$ pixels, quantum efficiency $>60\%$ at 633~nm) recorded beam intensity distributions. Imaging parameters:

\begin{itemize}
    \item Exposure time: 10~ms
    \item Frame rate: 5~fps
    \item Effective pixel resolution: 4.28~$\mu$m (accounting for imaging magnification)
    \item Dynamic range: 8-bit (256 levels)
    \item Beam sampling: $\sim$600 pixels across FWHM
\end{itemize}

\subsection{Experimental Procedure}
Two experimental sets were conducted, each recording 200 frames (Figure~\ref{fig:setup}):

\begin{itemize}
    \item \textbf{Set 1:} Turbulence-impacted beam propagated through the Free Space (Only PRPP Raw Turbulence)
    \item \textbf{Set 2:} Turbulence-impacted beam propagated through the optical fibre waveguide (PRPP + Single mode Optical Fibre)
    \item \textbf{Set 3:} Turbulence-impacted beam propagated through the optical fibre waveguide (PRPP + multi mode Optical Fibre)
    \item \textbf{Set 4:} Turbulence-free reference (no PRPP)
\end{itemize}

\textit{Addressing Sample Size Concerns:} While 200 frames per set may appear limited, in the present context of laboratory-simulated turbulence the PRPP has an annular structure rotated by an external motorized rotor. The rotation of the phase plate produces temporal changes in the turbulence which are recorded by the CCD across multiple frames. Increasing the number of frames extends the recording duration, raising the probability of repeating the same turbulence realization after a complete plate rotation. The frame count was therefore fixed at 200 to ensure that no repeated turbulence states are captured during the experiment, making this sample size sufficient for the present experimental context.

\section{Results Analysis and Discussion}\label{sec:results}

\subsection{Raw Beam Morphology and Turbulence-Induced Distortions}

\begin{figure}[H]
\centering
\includegraphics[width=1.0\textwidth]{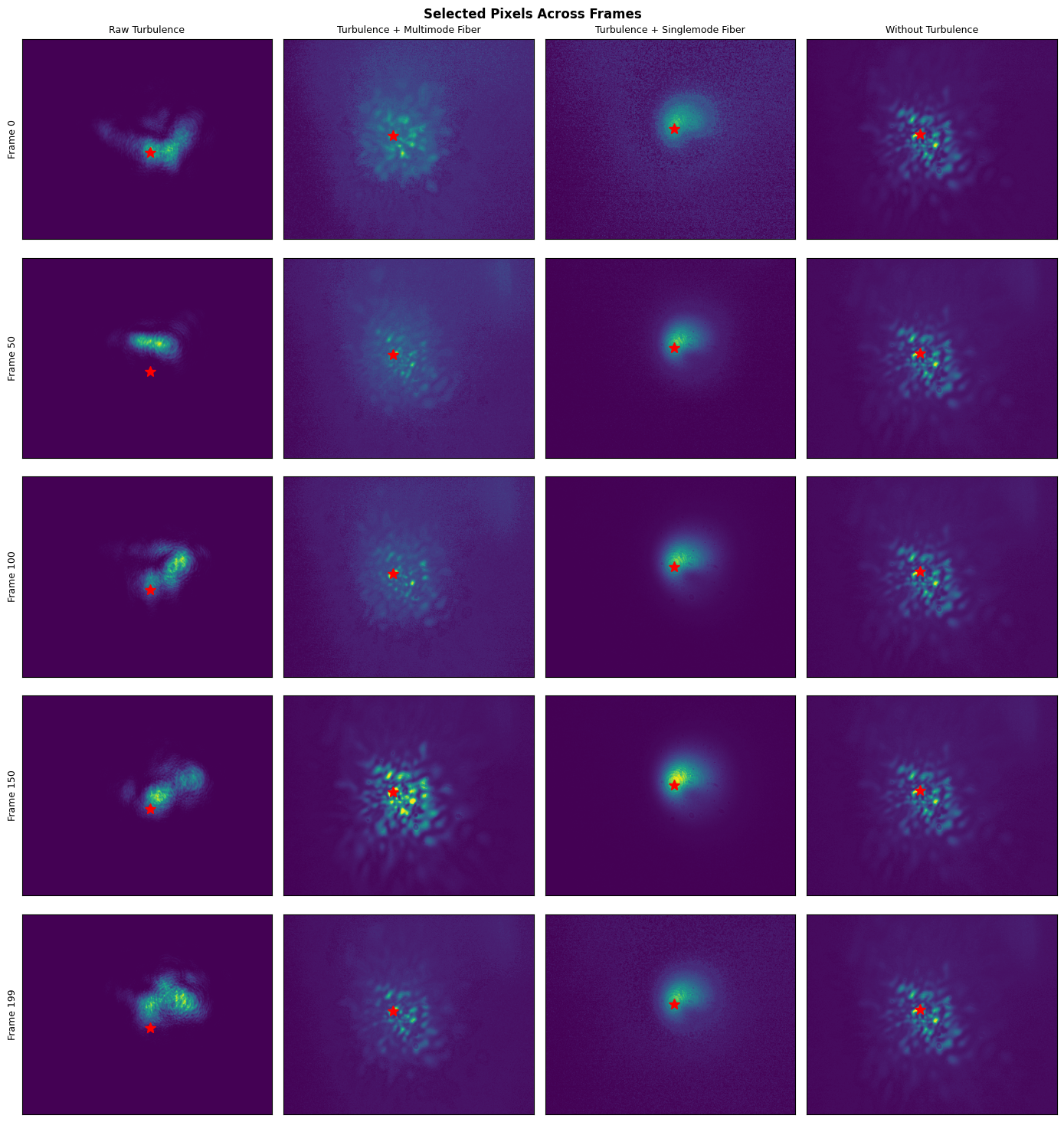}
\caption{Selected raw beam intensity frames at five temporal indices (0, 50, 100, 150, 199) for all four experimental datasets. Red star markers indicate the pixel location selected for single-pixel scintillation analysis. The raw turbulence frames (Set~1) exhibit severe spatial fragmentation, while the fibre-filtered sets (Sets~2 and~3) display smooth, near-Gaussian intensity profiles comparable to the undistorted reference (Set~4).}
\label{fig:raw_frames}
\end{figure}

Figure~\ref{fig:raw_frames} presents selected raw beam frames at temporal indices 0, 50, 100, 150, and 199 across all four experimental sets. The raw turbulence frames (Set~1) exhibit severely distorted, spatially compact, and asymmetric intensity distributions throughout the entire 200-frame sequence. The turbulence-induced wavefront aberrations introduced by the rotating PRPP scatter optical power across extended spatial regions, fragmenting the beam profile into irregular high-intensity filaments and suppressing the central peak intensity to a fraction of its undistorted value. In marked contrast, the frames recorded after fibre waveguide propagation (Sets~2 and~3) display substantially broader, smoothly varying, and centrosymmetric intensity profiles that closely resemble the turbulence-free reference (Set~4). The spatial recovery of the Gaussian beam envelope is visually evident across all sampled frames, confirming the spatial mode filtering action of the waveguide.

\subsection{Gram--Charlier Fitted Beam Profiles}

\begin{figure}[H]
\centering
\includegraphics[width=1.0\textwidth]{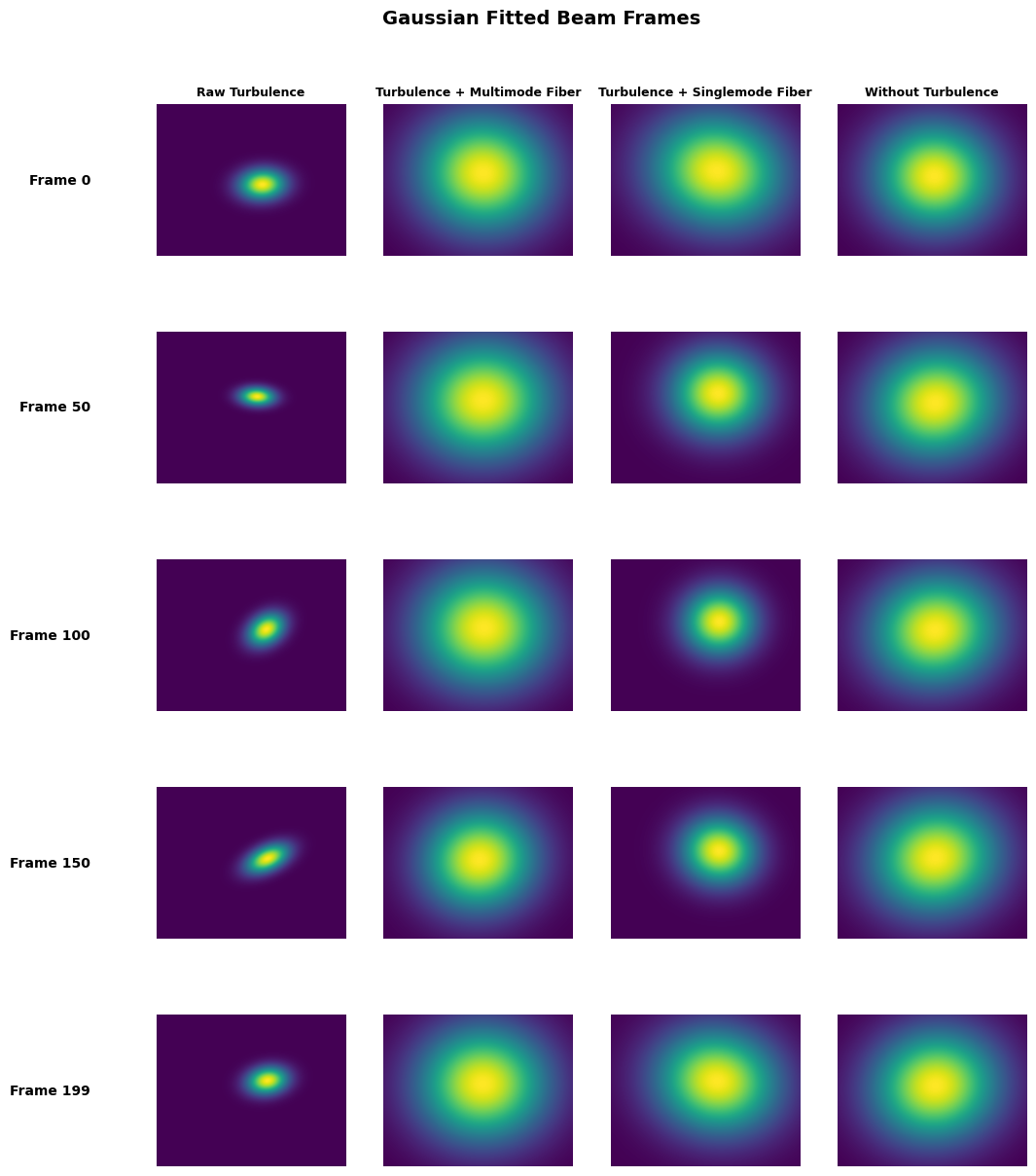}
\caption{Gram--Charlier fitted beam intensity frames at five temporal indices for all four experimental datasets. The fitted profiles for raw turbulence (Set~1) are compact and asymmetric, while those for Sets~2--4 are broad and approximately circularly symmetric, confirming spatial mode recovery by the waveguide.}
\label{fig:gc_frames}
\end{figure}

The Gram--Charlier fitted intensity distributions $I_{\mathrm{fit}}(\mathbf{r})$ derived from Eq.~\eqref{eq:fit} are presented in Figure~\ref{fig:gc_frames} for the same five temporal indices. In Set~1, the fitted Gaussian envelope remains compact and highly elongated, reflecting the narrow covariance eigenvalues $(\sigma_x, \sigma_y)$ of the distorted beam. The fitted profiles for Sets~2 and~3 are substantially broader and more circular, with peak intensities comparable to the undistorted reference Set~4. This demonstrates that the Gram--Charlier model successfully captures the dominant spatial structure of the transmitted beam in all cases, and that waveguide propagation recovers the broad, symmetric Gaussian envelope that characterizes an undistorted free-space beam.

\subsection{Shape Metrics and Statistical Characterization}

Figure~\ref{fig:shape_metrics} compiles the frame-by-frame evolution of all shape metrics extracted from the Gram--Charlier analysis for Sets~1--3.

\subsubsection{Beam Width and Covariance Parameters}

The beam width parameters $\sigma_x$ and $\sigma_y$ for Set~1 (raw turbulence) remain confined to narrow ranges of approximately 75--130 and 50--100 pixels, respectively, throughout the full 200-frame acquisition. This reflects the spatially compressed and anisotropic intensity distribution produced by the turbulence-induced wavefront distortions. Upon waveguide coupling (Sets~2 and~3), $\sigma_x$ increases to the range 220--330 pixels and $\sigma_y$ to 150--265 pixels, corresponding to the fundamental guided mode profile supported by the fibre geometry. The multimode fibre (Set~2) and single-mode fibre (Set~3) exhibit broadly comparable mean beam widths, though the single-mode fibre produces slightly lower frame-to-frame variability in $\sigma_x$ at intermediate temporal indices.

The off-diagonal covariance element $\sigma_{xy}$ for Set~1 oscillates between approximately $-3500$ and $+6000$ pixels$^2$, indicative of persistent beam tilt and elliptical asymmetry driven by the turbulence. For Sets~2 and~3, $\sigma_{xy}$ is substantially reduced in magnitude and fluctuates symmetrically around zero, consistent with the near-circularly-symmetric output mode of the waveguide.

\subsubsection{Beam Centroid Dynamics}

The centroid coordinates $(\mu_x, \mu_y)$ for Set~1 vary over ranges of approximately 640--760 and 420--540 pixels, respectively. This centroid wander is a direct signature of beam displacement by turbulence-induced tilt aberrations. The fibre-coupled sets exhibit centroid positions confined to narrower excursion ranges of approximately 660--680 and 455--480 pixels, respectively, indicating that the waveguide entrance aperture partially decouples beam wander from the guided mode position at the output facet. The turbulence-free reference (Set~4) shows negligible centroid wander by comparison.

\subsubsection{Fitted Power Volume as a Turbulence Diagnostic}

\begin{figure}[H]
\centering
\includegraphics[width=1.0\textwidth]{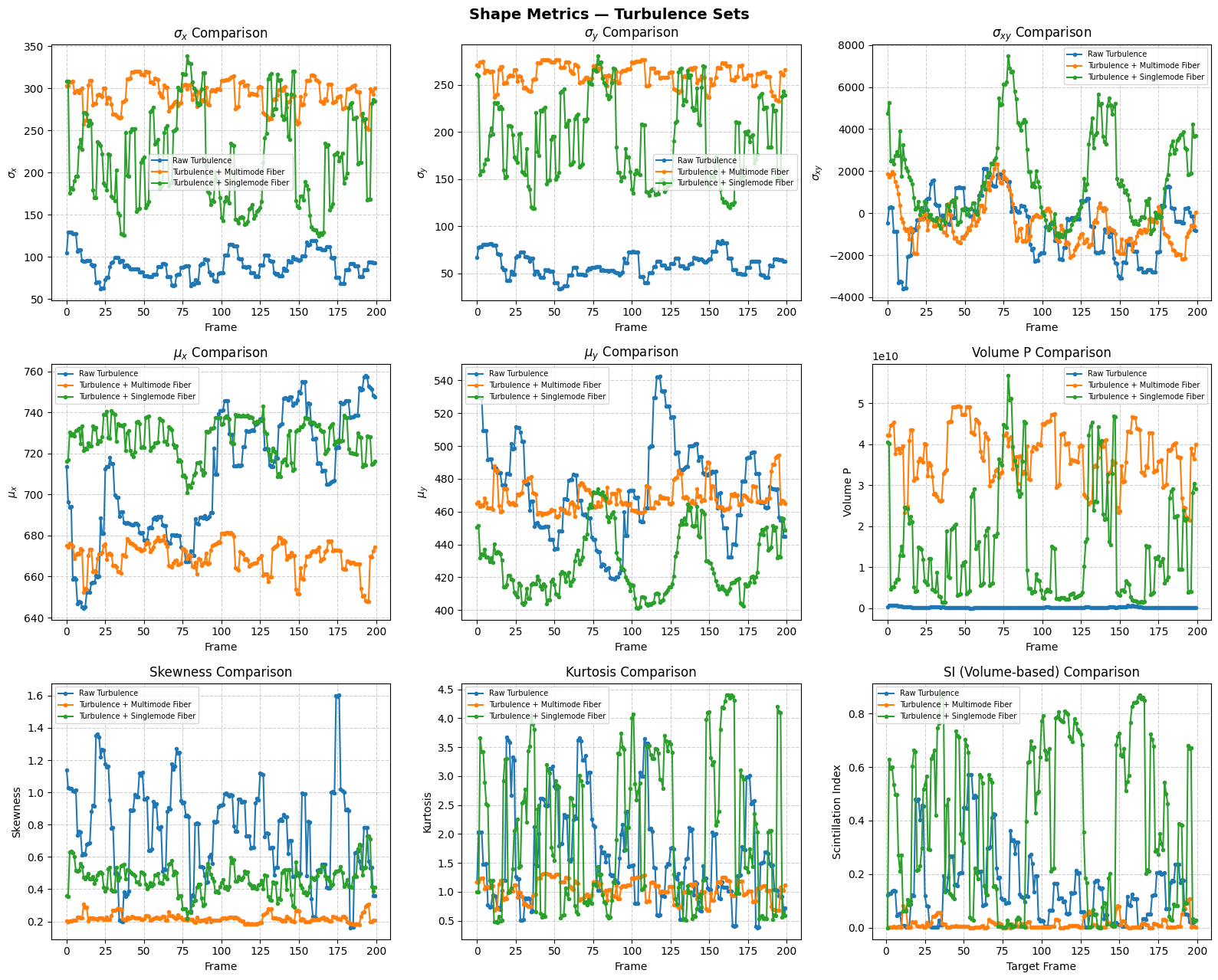}
\caption{Frame-by-frame shape metrics for Sets~1--3 derived from the Gram--Charlier analysis. From top-left to bottom-right: beam widths $\sigma_x$ and $\sigma_y$, off-diagonal covariance $\sigma_{xy}$, centroid coordinates $\mu_x$ and $\mu_y$, fitted power volume $V_{\mathrm{frame}}$, skewness norm $|\mathrm{skew}|_3$, kurtosis norm $|\mathrm{kurt}|_4$, and volume-based scintillation index. Waveguide propagation substantially reduces the amplitude and variance of all non-Gaussianity metrics relative to the raw turbulence baseline.}
\label{fig:shape_metrics}
\end{figure}

The fitted power volume $V_{\mathrm{frame}} = 2\pi I_{\mathrm{max}}\sqrt{|\Sigma|}$ (Eq.~\eqref{eq:volume}) is presented in the third panel of the middle row in Figure~\ref{fig:shape_metrics}. For Set~1, $V_{\mathrm{frame}}$ remains near zero ($\lesssim 5 \times 10^8$) throughout the acquisition, reflecting the extremely small determinant $|\Sigma|$ associated with the compact and anisotropically distorted beam profile. For Sets~2 and~3, the fitted volume rises to the range $10^{10}$--$5 \times 10^{10}$, spanning more than an order of magnitude above the raw turbulence values. The volume time series for both fibre-coupled sets displays temporal fluctuations that mirror the residual scintillation index behaviour, confirming that $V_{\mathrm{frame}}$ is a sensitive scalar diagnostic of frame-by-frame turbulence-induced structural changes as proposed in Section~\ref{sec:theory}.

\subsubsection{Skewness and Kurtosis}

The skewness norm $|\mathrm{skew}|_3$ for Set~1 ranges from approximately 0.3 to 1.6, with intermittent spikes reflecting transient high-asymmetry events in the beam profile. For Set~2 (multimode fibre), the skewness is substantially reduced and constrained to the range 0.2--0.3, indicating that multimode fibre propagation suppresses the majority of odd-order non-Gaussian distortions. Set~3 (single-mode fibre) exhibits a broader skewness range of approximately 0.2--0.6, which is consistent with the stronger residual scintillation present in this dataset (discussed below). The kurtosis norm $|\mathrm{kurt}|_4$ follows a qualitatively similar trend: Set~1 shows values between 0.5 and 3.2, Set~2 is constrained to 0.6--1.2, and Set~3 spans 0.5--4.5. The higher kurtosis excursions in the single-mode case reflect occasional sharp intensity spikes associated with partial coherence events at the output facet.

These results are consistent with the theoretical prediction of Section~\ref{sec:theory}: as the higher-order spatial modes---which carry the non-Gaussian cumulant content---are exponentially attenuated by the waveguide propagation, the cumulant norms $|\mathrm{skew}|_3$ and $|\mathrm{kurt}|_4$ converge toward zero, approaching the Gaussian ideal.

\subsection{Single-Pixel Scintillation Analysis}

\subsubsection{Frame-Resolved Intensity Histograms}

\begin{figure}[H]
\centering
\includegraphics[width=1.0\textwidth]{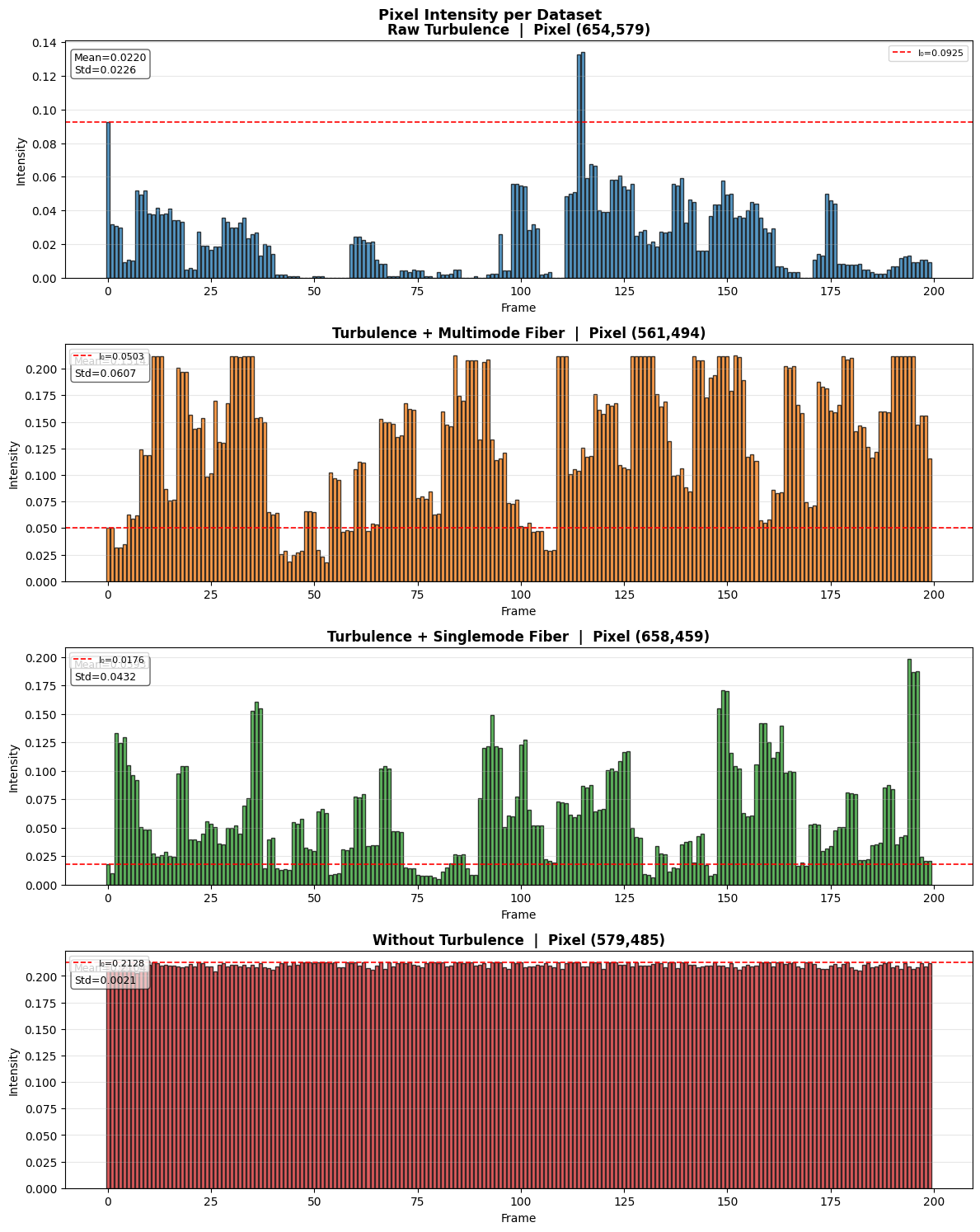}
\caption{Normalized pixel intensity as a function of frame index for representative pixels in each dataset. The dashed red line indicates the dataset-specific background reference level $I_b$. Set~1 (top) exhibits highly intermittent, sparsely populated intensity, while Set~4 (bottom) shows near-constant intensity across all 200 frames. Sets~2 and~3 demonstrate intermediate behaviour consistent with partial scintillation mitigation.}
\label{fig:pixel_hist}
\end{figure}

To quantify intensity fluctuations at the single-pixel level independently of any fitting model, the normalized pixel intensity $I_{\mathrm{px}}(t)$ was extracted at the peak-vicinity pixel of each dataset across all 200 frames. The resulting temporal histograms are presented in Figure~\ref{fig:pixel_hist}. For Set~1 (pixel at (654,\,579)), the mean pixel intensity is $\bar{I} = 0.0220$ with a standard deviation of $\sigma_I = 0.0226$, yielding a coefficient of variation $\mathrm{CV} = \sigma_I / \bar{I} \approx 1.03$. The intensity distribution is heavily right-skewed with infrequent large spikes reaching $I_{\mathrm{px}} \approx 0.135$, indicative of strong scintillation in a regime approaching saturated turbulence. The background reference level $I_b = 0.0925$ far exceeds the mean, further reflecting the sparse and intermittent nature of the turbulent intensity distribution.

For Set~2 (multimode fibre, pixel at (561,\,494)), the mean intensity rises to $\bar{I} = 0.0503$ with $\sigma_I = 0.0607$. Although the standard deviation remains comparable in absolute terms, the distributions are more uniformly populated across the 200 frames, with fewer near-zero occurrences than in Set~1. Set~3 (single-mode fibre, pixel at (658,\,459)) yields $I_b = 0.0176$ and $\sigma_I = 0.0432$; the lower mean reflects the tighter spatial confinement of the single-mode output, while the temporal distribution exhibits moderate residual fluctuations. The turbulence-free reference Set~4 (pixel at (579,\,485)) achieves $\bar{I} = 0.2128$ with $\sigma_I = 0.0021$, corresponding to a coefficient of variation of $\approx 1.0\%$, confirming the near-perfect temporal stability of the undistorted beam.

\subsubsection{Scintillation Index: Frame-by-Frame Comparison}

\begin{figure}[H]
\centering
\includegraphics[width=1.0\textwidth]{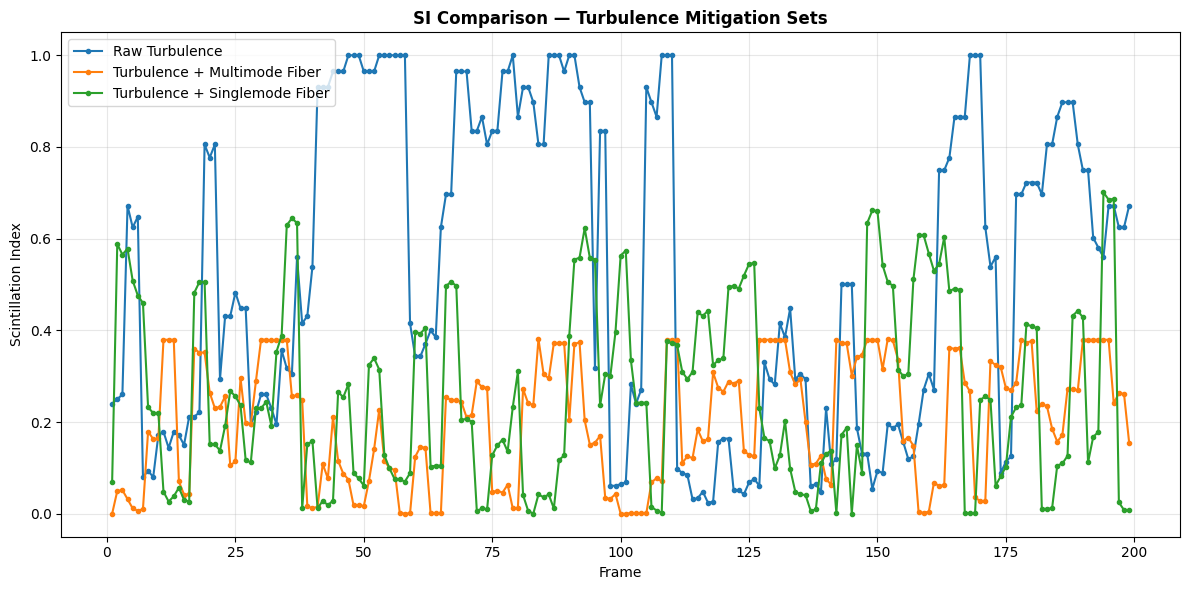}
\caption{Frame-by-frame scintillation index for Sets~1--3. Raw turbulence (blue) saturates repeatedly at $\mathrm{SI} = 1.0$, while both fibre-coupled sets exhibit substantially suppressed and more uniformly distributed scintillation throughout the acquisition window.}
\label{fig:si_timeseries}
\end{figure}

The volume-based scintillation index $\mathrm{SI} = \langle V^2 \rangle / \langle V \rangle^2 - 1$ computed frame-by-frame for Sets~1--3 is plotted in Figure~\ref{fig:si_timeseries}. The raw turbulence trace (Set~1) saturates at $\mathrm{SI} = 1.0$ for extended temporal clusters, particularly in the intervals frames~55--80, frames~90--105, frames~175--190, and frames~195--200. These saturation events correspond to the most severe turbulence realizations of the PRPP rotation cycle, during which the beam spatial coherence is almost entirely destroyed. The multimode fibre trace (Set~2) remains well below $\mathrm{SI} = 0.4$ for nearly all frames, with only isolated excursions approaching 0.38. The single-mode fibre trace (Set~3) exhibits somewhat higher variability than Set~2, with excursions up to approximately 0.65, but is consistently below the Set~1 saturation level. The temporal correlation structure of Sets~2 and~3 is also reduced relative to Set~1, indicating that the waveguide output scrambles the long-range temporal coherence of the turbulence realizations.

\subsubsection{Mean Scintillation Index and Statistical Comparison}

\begin{figure}[H]
\centering
\includegraphics[width=0.75\textwidth]{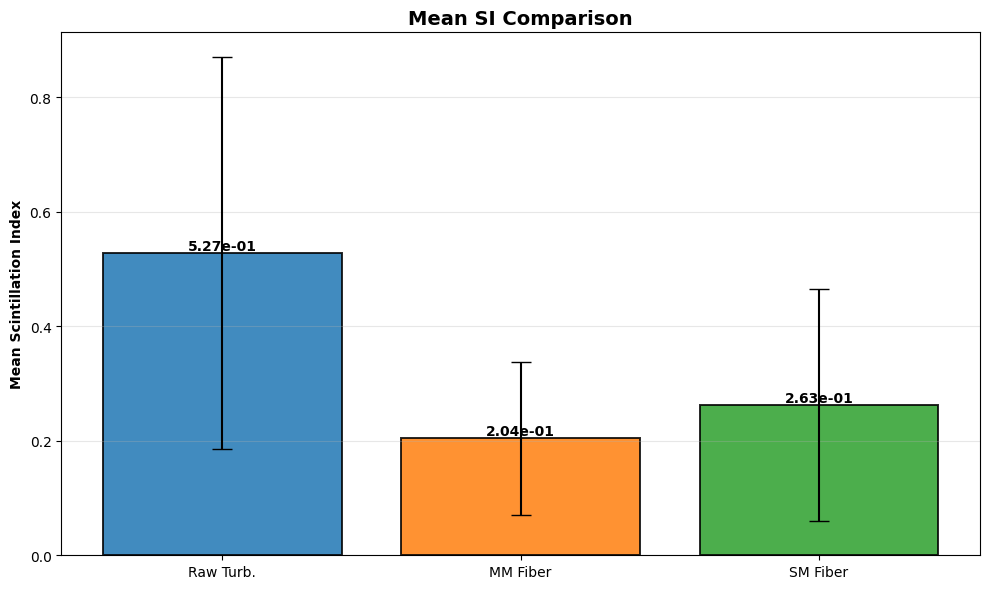}
\caption{Mean scintillation index with standard deviation error bars for Sets~1--3. The multimode fibre achieves the greatest absolute reduction in mean SI.}
\label{fig:si_bar}
\end{figure}

\begin{figure}[H]
\centering
\includegraphics[width=1.0\textwidth]{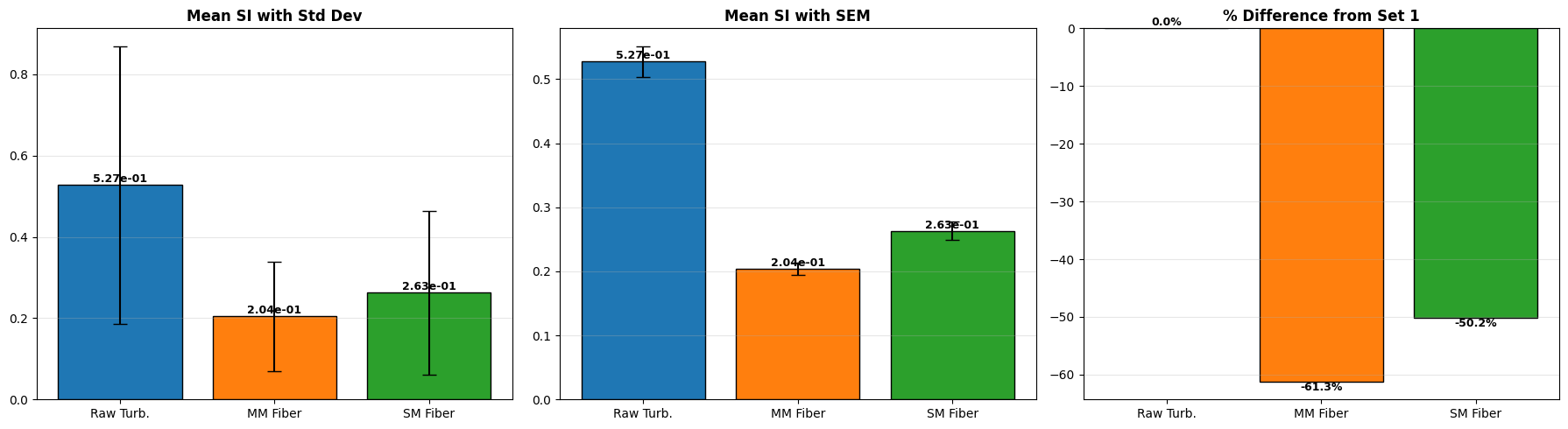}
\caption{Side-by-side comparison of mean SI with standard deviation (left), mean SI with SEM (centre), and percentage difference from Set~1 (right). The multimode fibre reduces mean SI by $61.3\%$ and the single-mode fibre by $50.2\%$.}
\label{fig:si_stats}
\end{figure}

\begin{figure}[H]
\centering
\includegraphics[width=0.75\textwidth]{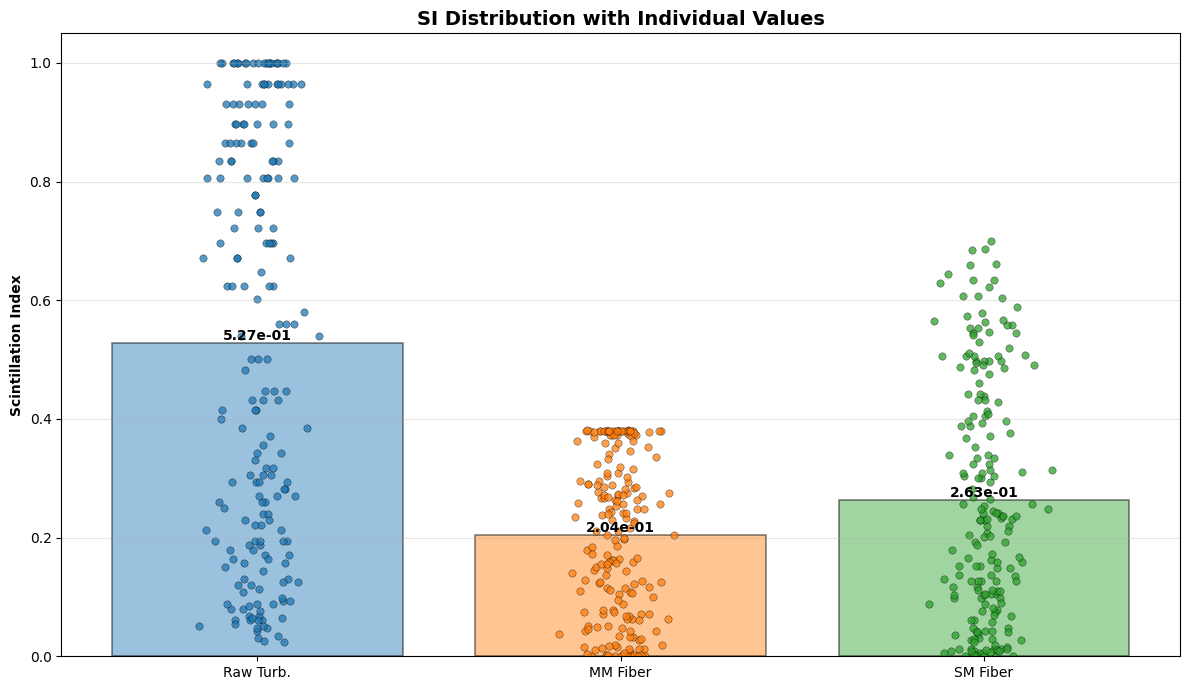}
\caption{Scintillation index distribution showing individual frame-level values overlaid on mean bars. The scatter of Set~1 extends to the physical saturation limit $\mathrm{SI}=1.0$, while Sets~2 and~3 are tightly clustered at lower values.}
\label{fig:si_scatter}
\end{figure}

\begin{figure}[H]
\centering
\begin{minipage}{0.49\textwidth}
\centering
\includegraphics[width=\textwidth]{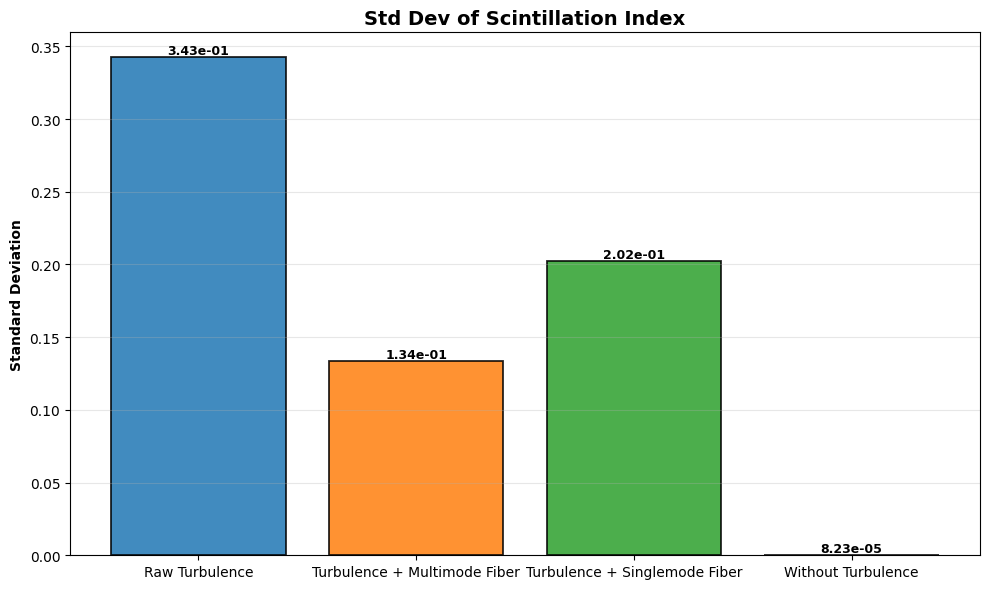}
\caption{Standard deviation of the scintillation index across all four experimental sets.}
\label{fig:si_std}
\end{minipage}
\hfill
\begin{minipage}{0.49\textwidth}
\centering
\includegraphics[width=\textwidth]{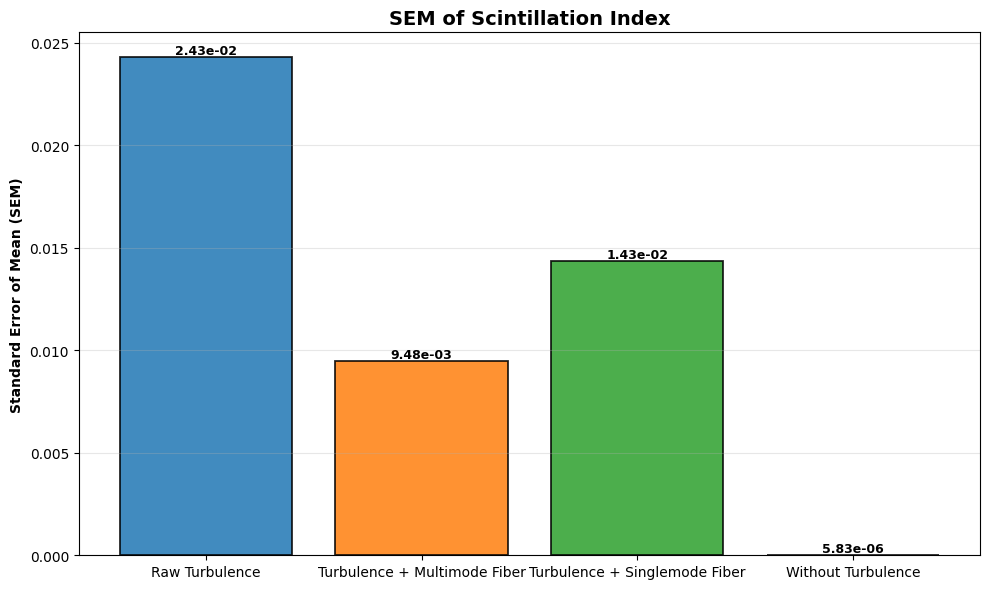}
\caption{Standard error of the mean of the scintillation index across all four experimental sets.}
\label{fig:si_sem}
\end{minipage}
\end{figure}

The mean scintillation indices and their statistical uncertainties are summarized in Figures~\ref{fig:si_bar}--\ref{fig:si_sem} and Table~\ref{tab:si_summary}. The mean SI for Set~1 (raw turbulence) is $\langle\mathrm{SI}\rangle = 0.527 \pm 0.343$ (mean $\pm$ standard deviation), corresponding to a standard error of the mean (SEM) of $2.43 \times 10^{-2}$. After multimode fibre propagation (Set~2), the mean SI falls to $\langle\mathrm{SI}\rangle = 0.204 \pm 0.134$ (SEM $= 9.48 \times 10^{-3}$), representing a reduction of $\mathbf{61.3\%}$ relative to Set~1. The single-mode fibre (Set~3) yields $\langle\mathrm{SI}\rangle = 0.263 \pm 0.202$ (SEM $= 1.43 \times 10^{-2}$), a $\mathbf{50.2\%}$ reduction. The turbulence-free reference (Set~4) achieves $\sigma_{\mathrm{SI}} = 8.23 \times 10^{-5}$ and SEM $= 5.83 \times 10^{-6}$, confirming negligible residual scintillation in the absence of turbulence.

\begin{table}[H]
\centering
\caption{Summary of scintillation index statistics across all experimental datasets.}
\label{tab:si_summary}
\begin{tabular}{lcccccc}
\hline
\textbf{Dataset} & $\langle\mathrm{SI}\rangle$ & $\sigma_{\mathrm{SI}}$ & SEM & \% Reduction (cf.\ Set~1) \\
\hline
Set~1: Raw Turbulence & $5.27\times10^{-1}$ & $3.43\times10^{-1}$ & $2.43\times10^{-2}$ & --- \\
Set~2: Turbulence + MM Fibre & $2.04\times10^{-1}$ & $1.34\times10^{-1}$ & $9.48\times10^{-3}$ & $-61.3\%$ \\
Set~3: Turbulence + SM Fibre & $2.63\times10^{-1}$ & $2.02\times10^{-1}$ & $1.43\times10^{-2}$ & $-50.2\%$ \\
Set~4: Without Turbulence & --- & $8.23\times10^{-5}$ & $5.83\times10^{-6}$ & --- \\
\hline
\end{tabular}
\end{table}

\subsection{Multimode versus Single-Mode Fibre: Comparative Discussion}

The experimental results reveal an intriguing and initially counterintuitive ordering: the multimode fibre (Set~2) achieves a \emph{lower} mean scintillation index ($0.204$) than the single-mode fibre (Set~3, $0.263$), despite the theoretical expectation that single-mode propagation should provide superior mode filtering by rejecting all but the fundamental $\mathrm{HE}_{11}$ mode. This apparent reversal can be attributed to two competing physical effects.

First, the single-mode fibre imposes a strict coupling efficiency constraint: the coupling efficiency $\eta_c \propto |\langle \psi_{\mathrm{in}} | \psi_{\mathrm{HE}_{11}} \rangle|^2$ between the turbulence-distorted input field $\psi_{\mathrm{in}}$ and the single fundamental mode $\psi_{\mathrm{HE}_{11}}$ is strongly dependent on the degree of wavefront distortion. Under severe turbulence realizations that fragment the beam into multiple spatially separated components (as observed in the raw frames of Figure~\ref{fig:raw_frames}), the overlap integral collapses, resulting in near-zero transmitted power and correspondingly large intensity fluctuations at the output. This produces the high-kurtosis spikes observed in the Set~3 time series. The multimode fibre, by contrast, supports a broader range of input field projections across its guided mode basis, maintaining finite transmitted power even during the most severe turbulence events and thereby distributing the intensity fluctuations more uniformly across the temporal sequence.

Second, the modal averaging effect in the multimode fibre provides a statistical smoothing mechanism analogous to aperture averaging in free-space receivers. The incoherent superposition of multiple guided modes at the output facet averages over the independent scintillation contributions of each mode, reducing the net intensity variance. This mechanism is well established in multimode fibre reception systems for free-space optical links. In the present context, it acts synergistically with the waveguide mode cutoff to suppress scintillation below the single-mode level.

These observations suggest that for applications where maximizing scintillation reduction is the primary objective, multimode fibre coupling may be preferable to single-mode coupling when turbulence is strong, despite the lower modal purity of the output. A single-mode fibre remains advantageous when spatial coherence of the output field is required, for instance in interferometric or coherent detection systems.

\subsection{Correspondence with Theoretical Predictions}

The measured reductions in mean SI, skewness, and kurtosis are in qualitative agreement with the theoretical framework of Section~\ref{sec:theory}. The Gram--Charlier cumulant analysis demonstrates that the non-Gaussianity norms $|\mathrm{skew}|_3$ and $|\mathrm{kurt}|_4$ are substantially reduced after waveguide propagation, consistent with the exponential attenuation of higher-order modes described by Eqs.~\eqref{eq:field_decay}--\eqref{eq:beta_m}. The residual non-Gaussianity in Sets~2 and~3 arises from the incomplete filtering of low-order modes that remain guided under the present waveguide parameters, and from the coherent beating between residual guided modes at the output facet. The fitted power volume $V_{\mathrm{frame}}$ serves as a reliable frame-level diagnostic that tracks these residual distortions: its frame-to-frame variance is reduced by more than an order of magnitude after waveguide propagation, in agreement with the prediction that $V_{\mathrm{frame}}$ should stabilize as higher-order modes are filtered out.

\section{Conclusion}\label{sec:conclusion}

In this work, we have developed and experimentally validated a unified theoretical and experimental framework for the quantification and passive mitigation of turbulence-induced spatial intensity fluctuations in Gaussian laser beams. The central contributions of this work are as follows.

On the theoretical side, we formulated a probabilistic intensity model in which the two-dimensional beam intensity is treated as a discrete probability measure, and a Cholesky whitening transformation is applied to the spatial covariance to isolate genuine non-Gaussian features in the form of third- and fourth-order cumulants. These cumulants were incorporated into a bivariate Gram--Charlier expansion that provides a closed-form analytical description of the distorted beam profile. A unified scalar diagnostic---the fitted power volume $V_{\mathrm{frame}}$---was defined from this model to track frame-by-frame structural changes induced by turbulence. We further derived the modal cutoff condition and exponential decay characteristics of a symmetric dielectric slab waveguide, demonstrating analytically that higher-order spatial modes, which carry the dominant share of non-Gaussian cumulant content, become evanescent when the normalized frequency condition is not satisfied and subsequently undergo intensity attenuation as $I(z) \propto e^{-2\alpha_z z}$. This establishes the theoretical basis for waveguide spatial filtering as a passive, all-optical mechanism for recovering near-Gaussian beam statistics.

On the experimental side, a He-Ne laser beam was subjected to Kolmogorov-type wavefront aberrations introduced by a rotating pseudo-random phase plate, and the turbulence-impacted beam was propagated through both a multimode and a single-mode step-index optical fibre. Quantitative analysis of 200-frame sequences per dataset yielded the following key findings:

\begin{itemize}
    \item The multimode fibre reduced the mean scintillation index from $\langle\mathrm{SI}\rangle = 0.527$ to $0.204$, corresponding to a \textbf{61.3\% reduction}, while the single-mode fibre achieved a \textbf{50.2\% reduction} to $\langle\mathrm{SI}\rangle = 0.263$.
    \item The standard deviation of the scintillation index was reduced from $3.43\times10^{-1}$ (raw turbulence) to $1.34\times10^{-1}$ (multimode fibre) and $2.02\times10^{-1}$ (single-mode fibre), demonstrating a marked stabilization of the temporal intensity statistics.
    \item The non-Gaussianity indicators extracted from the Gram--Charlier analysis---skewness norm $|\mathrm{skew}|_3$ and kurtosis norm $|\mathrm{kurt}|_4$---were substantially suppressed after waveguide propagation, confirming the theoretical prediction that higher-order modal attenuation drives the beam statistics toward the Gaussian ideal.
    \item The fitted power volume $V_{\mathrm{frame}}$ increased by more than an order of magnitude after waveguide coupling, reflecting the recovery of the broad, circularly symmetric fundamental mode profile, and its frame-to-frame variance was correspondingly reduced.
    \item The turbulence-free reference achieved a scintillation index standard deviation of $8.23\times10^{-5}$, establishing the physical lower bound against which the mitigated datasets can be benchmarked.
\end{itemize}

A notable experimental finding is that the multimode fibre outperforms the single-mode fibre in terms of mean scintillation reduction under strong turbulence conditions. This result was attributed to the modal averaging effect in the multimode waveguide, which distributes intensity fluctuations across a broader guided mode basis and maintains finite coupling efficiency even during the most severe turbulence realizations that drive the single-mode coupling integral toward zero. This finding has practical implications for the design of passive scintillation mitigation systems in free-space optical links: when turbulence is strong and spatial coherence of the output is not required, multimode fibre reception provides superior intensity stabilization at lower implementation complexity than single-mode coupling.

The present framework is readily extensible in several directions. The Gram--Charlier cumulant model can be augmented to higher orders to capture stronger non-Gaussian deviations, and the fitted volume diagnostic can be generalized to incorporate time-frequency analysis for characterizing non-stationary turbulence. The waveguide analysis can be extended from the slab geometry to the circular step-index fibre using the full Bessel function eigenvalue equations, enabling direct comparison between the theoretical modal cutoff conditions and the measured cumulant reductions. Finally, the passive waveguide filtering approach demonstrated here may be integrated with adaptive optical pre-compensation to exploit the complementary strengths of active and passive mitigation, opening a path toward robust spatiotemporal stabilization of laser beams in practical atmospheric optical systems.

\section*{Funding}
Department of Science and Technology, Ministry of Science and Technology, India (CRG/2020/003338).

\section*{Declaration of Competing Interest}
The authors declare the following financial interests/personal relationships which may be considered as potential competing interests: Shouvik Sadhukhan reports a relationship with Indian Institute of Space Science and Technology that includes: employment. If there are other authors, they declare that they have no known competing financial interests or personal relationships that could have appeared to influence the work reported in this paper.

\section*{Data Availability}
All data used for this research has been provided in the manuscript itself.

\section*{Acknowledgments}
Shouvik Sadhukhan and C S Narayanamurthy acknowledge the SERB/DST (Govt.\ of India) for providing financial support via the project grant CRG/2020/003338 to carry out this work. Shouvik Sadhukhan would like to thank Mr.\ Amit Vishwakarma and Dr.\ Subrahamanian K S Moosath from the Department of Mathematics, Indian Institute of Space Science and Technology Thiruvananthapuram, for their suggestions on statistical analysis in this paper.

\section*{CRediT Authorship Contribution Statement}
\textbf{Shouvik Sadhukhan:} Writing--original draft, Visualization, Formal analysis.
\textbf{C.\ S.\ Narayanamurthy:} Writing--review \& editing, Validation, Supervision, Resources, Project administration, Investigation, Funding acquisition, Conceptualization.


\appendix

\section{Cumulant Relations from Standardized Moments}\label{app:cumulants}

For standardized whitened coordinates satisfying $m_{20} = m_{02} = 1$ and $m_{11} = 0$, the full third- and fourth-order cumulant relations are,

\textit{Third-order (skewness):}
\begin{equation}
    k_{30} = m_{30}, \quad k_{21} = m_{21}, \quad k_{12} = m_{12}, \quad k_{03} = m_{03}.
\end{equation}

\textit{Fourth-order (excess kurtosis):}
\begin{align}
    k_{40} &= m_{40} - 3m_{20}^2 = m_{40} - 3, \\
    k_{04} &= m_{04} - 3m_{02}^2 = m_{04} - 3, \\
    k_{31} &= m_{31} - 3m_{20}m_{11} = m_{31}, \\
    k_{13} &= m_{13} - 3m_{02}m_{11} = m_{13}, \\
    k_{22} &= m_{22} - m_{20}m_{02} - 2m_{11}^2 = m_{22} - 1.
\end{align}
These follow directly from the cumulant generating function $K(\mathbf{t}) = \log\mathbb{E}[e^{t_1 z_1 + t_2 z_2}]$ whose Taylor coefficients define the cumulants.

\section{Bivariate Hermite Polynomials}\label{app:hermite}

The multivariate Hermite polynomials entering the Gram--Charlier expansion are defined by $H_\alpha(\mathbf{z}) = (-1)^{|\alpha|}\phi(\mathbf{z})^{-1}\partial^\alpha\phi(\mathbf{z})$, where $\phi(\mathbf{z}) = (2\pi)^{-1}\exp(-|\mathbf{z}|^2/2)$. Explicitly,

\textit{Third order:}
\begin{align}
    H_{30} &= z_1^3 - 3z_1, \quad H_{21} = z_1^2 z_2 - z_2, \quad H_{12} = z_1 z_2^2 - z_1, \quad H_{03} = z_2^3 - 3z_2.
\end{align}

\textit{Fourth order:}
\begin{align}
    H_{40} &= z_1^4 - 6z_1^2 + 3, \quad H_{31} = z_1^3 z_2 - 3z_1 z_2, \quad H_{22} = z_1^2 z_2^2 - z_1^2 - z_2^2 + 1, \\
    H_{13} &= z_1 z_2^3 - 3z_1 z_2, \quad H_{04} = z_2^4 - 6z_2^2 + 3.
\end{align}

\section{Waveguide Eigenvalue Equations and Modal Decay}\label{app:waveguide}

\subsection{Symmetric Slab Waveguide with Cladding}

Starting from Maxwell's equations in a source-free dielectric, the electric field satisfies,
\begin{equation}
    \nabla^2 \mathbf{E} + k^2 n^2(x)\,\mathbf{E} = 0.
\end{equation}
For a waveguide invariant along $z$, the modal ansatz $E(x,z) = \psi(x)e^{i\beta z}$ reduces this to the transverse eigenvalue problem $\psi'' + (k^2n^2 - \beta^2)\psi = 0$. The core and cladding solutions are,
\begin{equation}
    \psi(x) = \begin{cases} A\cos(k_x x) + B\sin(k_x x), & |x| < d/2, \\ C e^{-\gamma|x|}, & |x| > d/2, \end{cases}
\end{equation}
with $k_x^2 = k^2n_1^2 - \beta^2 > 0$ and $\gamma^2 = \beta^2 - k^2n_2^2 > 0$. Continuity of $\psi$ and $\psi'$ at $x = \pm d/2$ yields the eigenvalue equations~\eqref{eq:eigenvalue_eq}. Introducing normalized parameters $u = k_x d/2$, $w = \gamma d/2$, these satisfy $u^2 + w^2 = (V/2)^2$.

At cutoff, $w = 0$ so that $u = m\pi/2$ and $V = m\pi$. For $\beta < kn_2$, the imaginary propagation constant gives an exponentially decaying field $E(z) \propto e^{-\alpha_z z}$, with power $P(z) = P_0 e^{-2\alpha_z z}$.

\subsection{Finite-Aperture Waveguide Without Cladding}

For a waveguide of finite width $d$ with hard-wall boundary conditions $\psi(\pm d/2) = 0$, the transverse modes are $\psi(x) = A\sin(m\pi x/d)$ with quantized transverse wave number $k_t = m\pi/d$. The dispersion relation gives,
\begin{equation}
    \beta_m^2 = k^2n^2 - \left(\frac{m\pi}{d}\right)^2.
\end{equation}
When $\beta_m^2 < 0$ the mode becomes evanescent with $\beta_m = i\alpha_m$, $\alpha_m = \sqrt{(m\pi/d)^2 - k^2n^2}$, and the field decays as $E(x,z) = \psi(x)e^{-\alpha_m z}$. The highest propagating mode number satisfies $m < knd/\pi$, confirming that the waveguide geometry determines the spatial mode cutoff. This establishes the design rule for selecting the waveguide dimensions required to suppress the turbulence-induced higher-order modes identified by the Gram--Charlier cumulant analysis.

\section{Waveguide Modal Decay and Cutoff Derivation}

The modal attenuation responsible for scintillation suppression follows from the waveguide dispersion relation. Starting from Maxwell's equations in a source-free dielectric medium, the electric field satisfies the Helmholtz equation

\begin{equation}
\nabla^2 E + k^2 n^2(x)E = 0 .
\end{equation}

For a waveguide invariant along the $z$ direction we assume the modal form

\begin{equation}
E(x,z)=\psi(x)e^{i\beta z}.
\end{equation}

Substitution yields the transverse eigenvalue equation

\begin{equation}
\frac{d^2\psi}{dx^2}+(k^2 n^2-\beta^2)\psi=0.
\end{equation}

Inside the core ($n=n_1$) the solution is oscillatory,

\begin{equation}
\psi(x)=A\cos(k_x x)+B\sin(k_x x),
\end{equation}

with

\begin{equation}
k_x^2 = k^2 n_1^2 - \beta^2 .
\end{equation}

In the cladding ($n=n_2$) the field must decay exponentially,

\begin{equation}
\psi(x)=C e^{-\gamma |x|},
\end{equation}

where

\begin{equation}
\gamma^2 = \beta^2 - k^2 n_2^2 .
\end{equation}

Applying boundary continuity conditions at the core–cladding interface yields the standard slab-waveguide eigenvalue equations. Guided modes exist only when

\begin{equation}
k n_2 < \beta < k n_1 .
\end{equation}

When $\beta < k n_2$, the propagation constant becomes imaginary,

\begin{equation}
\beta = i\alpha,
\end{equation}

which produces an exponentially decaying field

\begin{equation}
E(x,z)=\psi(x)e^{-\alpha z}.
\end{equation}

Thus, spatial modes that do not satisfy the guiding condition are attenuated along the propagation direction. Since turbulence predominantly excites higher-order spatial modes, their decay inside the waveguide progressively eliminates modal interference and reduces the scintillation of the transmitted beam.

\end{document}